\documentstyle[12pt,epsfig]{article}
\def\be {\begin{equation}}
\def\ee {\end{equation}}
\def\bea {\begin{eqnarray}}
\def\eea {\end{eqnarray}}

\begin{document}

\thispagestyle{empty}
\vskip 15pt

\begin{center}
{\Large {\bf Gluon dissociation of $J/\psi$ in anisotropic 
{\em Quark-Gluon-Plasma}}}
\renewcommand{\thefootnote}{\alph{footnote}}

\hspace*{\fill}

\hspace*{\fill}

{ \tt{
Mahatsab Mandal\footnote{E-mail address:
mahatsab.mandal@saha.ac.in}
and
Pradip Roy\footnote{E-mail address:
pradipk.roy@saha.ac.in}
}}\\

\small {\em Saha Institute of Nuclear Physics \\
1/AF Bidhannagar, Kolkata - 700064, INDIA}
\\

\vskip 40pt

{\bf ABSTRACT}
\end{center}

\vskip 0.5cm

We calculate the gluon dissociation 
cross-section in an anisotropic quark gluon plasma expected to be formed in
relativistic nucleus-nucleus collisions. It is shown that the thermally
weighted cross-section of gluon dissociation undergoes 
modification in anisotropic plasma affecting the $J/\psi$
survival probability.
The dependence of the cross section on the direction of propagation 
of the charmonium with respect to the
anisotropy axis is presented. 
Survival probability of $J/\psi$ in two different space time
models of anisotropic quark gluon plasma (AQGP) has been calculated.
It is shown that depending upon the initial conditions (corresponding
to RHIC energies), the survival
probability in AQGP differs from that in isotropic QGP both in the
central as well as forward rapidity regions. 
For initial conditions relevant for LHC energies, marginal difference
between the the two space time models has been observed with a given initial
conditions.
\vskip 30pt

\section{Introduction}
Ever since the possibility of creating quark gluon plasma (QGP)
in relativistic heavy ion collision was envisaged, numerous
signals were proposed to probe the properties of such an exotic
state of matter.  In this context
Satz and Matsui~\cite{satz} had suggested that
the production of heavy quark resonances ($J/\psi$) will be suppressed
as a result of colour Debye screening in a hot
and dense system of quarks, anti-quarks and gluons.
This suppression could
be detected experimentally through the dileptonic decay mode of these
resonances. ALICE dimuon spectrometer~\cite{muonlab} is dedicated to look
for this
type of signal. However, it is a daunting task to
disentangle  the contributions of the heavy quarkonium states
to muon spectrum due to the
background from several other sources, {\it e.g.}
Drell-Yan, semileptonic
decay of open heavy flavoured mesons ($D{\bar D}, B{\bar B}$) etc.
Low energy muons
from kaons and pions also constitute a large background.

In a QGP the much harder gluons can easily break up a $J/\psi$ contrary
to the case of hadronic system. In equilibrating plasma the gluons have much
harder momentum sufficient to dissociate the charmonium. Such a study has been
performed in Ref.~\cite{xu} quite some time ago. Ever since the physical
picture of quarkonium dissociation has undergone slight refinement during the
couple of years. First of all, most of the existing calculations
on various observables assume, from the
very beginning, that the plasma is isotropic which may not necessarily be true
as we shall argue in the following. Moreover, many properties of the QGP
are still poorly understood. The measurement of elliptic
flow parameter and its theoretical explanation suggest that the matter
quickly comes into thermal equilibrium
(with $\tau_{\rm therm} < $ $1$ fm/c, where $\tau_{therm}$ is the time of
thermalization)~\cite{PRC75_ref1}.
As for example,
one of the major difficulty is to measure the thermalization
($\tau_{therm}$) and isotropization ($\tau_{iso}$) time of the QGP. On
the one hand, the success of ideal hydrodynamic fits to experimental
data~\cite{PRC75_ref1} implies rapid thermalization of the bulk matter
created at RHIC.
On the contrary, perturbative estimation suggests relatively
slower thermalization of
QGP~\cite{PRC75_ref2}. However, recent hydrodynamical
studies~\cite{0805.4552_ref4} have shown that
due to the poor knowledge of the initial conditions there is a
sizable amount of uncertainty in the estimate of thermalization or
isotropization time. It is suggested that (momentum) anisotropy driven
plasma instabilities may speed up the process of
isotropization~\cite{plb1181993}, in that case one is allowed to use
hydrodynamics for the evolution of the matter. However, instability-driven
isotropization is not yet proved at RHIC and LHC energies.

In absence of a theoretical proof favoring the rapid
thermalization and the uncertainties in the hydrodynamical fits of
experimental data, it is very hard to assume hydrodynamical
behavior of the system from the very beginning. 
The rapid expansion of the matter along the beam direction causes 
faster cooling in the longitudinal direction than in the transverse
direction~\cite{PRC75_ref2}. As a result, the system becomes anisotropic
with $\langle{p_L}^2\rangle << \langle{p_T}^2\rangle$ in the local rest frame.
At some later time when the effect of parton interaction rate
overcomes the plasma expansion rate, the system returns to the
isotropic state again and remains isotropic for the rest of the period.
Therefore, it has been suggested to look for some observables which 
are sensitive to the early time after the collision. The effects of
pre-equilibrium momentum anisotropy on various observables have been
studied quite extensively over the past few years. Heavy quark energy loss
and momentum broadening in anisotropic QGP have been studied in 
Refs.~\cite{12ofplb,13ofplb}. Effects of anisotropy on photon and
dilepton yields have been investigated rigorously in 
Ref.~\cite{LB1,LB2,LB3,schenke,dilstrick}. The effect of
initial state momentum anisotropy on the
radiative energy loss has been demonstrated in Ref.~\cite{prabhee}.
Recently, the authors in
Ref.~\cite{mahatsab}
calculated the nuclear modification factor for light hadrons assuming an 
anisotropic QGP and showed how the isotropization time can be
extracted by comparing with the experimental data. 
Most importantly, 
the heavy quark potential has been calculated in Ref.~\cite{18ofplb} 
and the solutions of Schrodinger equations have been obtained
in Ref.~\cite{19ofplb} corresponding to anisotropic system.  

It is to be noted that the calculations of $J/\psi$ dissociation
cross-section in Ref.~\cite{xu}
have been performed in an equilibrating plasma and it is
found that the survival probability increases in such system.
We, in the present work, shall extend the above work assuming initial
state momentum space anisotropy.

The plan of the paper is the following. In section 2 we briefly
recall the necessary ingredients to calculate the thermally weighted
gluon dissociation cross section in
anisotropic media. Then we discuss how this can be implemented to
calculate the survival probability of $J/\psi$ along with space-time models for
the anisotropic media.
Section 3 will be devoted
to discuss the results. Finally, we conclude in section 4.

\section{Formalism}
\subsection{The thermal-averaged Gluon-$J/\psi$ dissociation cross section}
Peskin and Bhanot first calculated the quarkonium-hadron interaction 
cross section using operator product expansion~\cite{peskin}. Similar
result was obtained using the QCD factorization theorem in Ref.~\cite{prc65}.
Same formalism allows to express the hadron-$J/\psi$ inelastic cross 
section in terms of the convolution of the inelastic gluon-$J/\psi$ 
dissociation cross section with the gluon distribution inside the 
hadron.The perturbative prediction for gluon-$J/\psi$ dissociation cross 
section is \cite{KSH}
\begin{equation}
 \sigma (q^0)= \frac{2\pi}{3}\left(\frac{32}{3}\right)^2
  \left(\frac{16\pi}{3g_s^2}\right)\frac{1}{m^2_Q}
  \frac{(q^0/\epsilon_0-1)^{3/2}}{(q^0/\epsilon_0)^5}\; , \label{eq1}
\end{equation}
where $q^0$ the energy of the gluon in the stationary $J/\psi$ frame;
$\epsilon_0$ is the binding energy of the $J/\psi$ where
$q_0 > \epsilon_0$. $g_s$ is the coupling constant and $m_Q$ is charm quark 
mass. A few comments about the binding energy of  quarkonium states is
in order here. It is to be noted that We have used the constant binding 
energy of
the $J/\psi$ in AQGP at finite temperature. However, using
the real and imaginary part of heavy quark potential (calculated
in anisotropic QGP) in Schrodinger equation the authors of 
Ref.~\cite{striclandprl&prd} have shown that the binding energy
of quarkonium states strongly depends on the anisotropy parameter
as well as on the hard momentum scale. This observation might
have important consequences on the gluon dissociation cross-section and 
hence on the survival probability. 

we assume that the J/$\psi$ moves with four-momentum $P$ given by
\begin{equation}
P=(M_T\cosh y,0,P_T,M_T\sinh y) \label{eq2}
\end{equation}
where $M_T=\sqrt{M_{J/\psi}^2+P_T^2}$ is the $J/\psi$ transverse mass 
and $y$ is the rapidity of the $J/\psi$. A gluon with a 
four-momentum $K=(k^0,{\bf k})$ in the rest frame of the parton gas has 
energy $q^0=K.u$ in the rest frame of the $J/\psi$. 
Now to calculate the velocity averaged cross section in anisotropic media
we note that the anisotropicity enters through the distribution function
\cite{LB3,dilstrick},
\begin{equation}
f(k^0,\xi,p_{\rm hard})=\frac{1}{e^{k^0/p_{\rm hard}\sqrt{1+\xi(\hat{k}.\hat{n})^2}}-1} \label{eq4}
\end{equation}
where $p_{\rm hard}$ is the hard momentum scale, $\hat{n}$ is the 
direction of anisotropy which is along the beam axis
and the parameter $\xi$ is the  
anisotropy parameter$(-1<\xi<\infty)$. $p_{\rm hard}$ is related to the average 
momentum in the partonic distribution function. 
In isotropic case, $\xi=0$ and $p_{\rm hard}$ can be identified with the 
temperature.
In such case the gluon-$J/\psi$ dissociation cross section becomes \cite{xu}, 
\begin{equation}
\langle\sigma(K.u) v_{\rm rel}\rangle_k=
\frac{\int d^3k \sigma(K.u)v_{\rm rel}f(k^0,\xi,p_{\rm hard})}{\int d^3kf(k^0,\xi,p_{\rm hard})} \label{eq3}
\end{equation}
where $v_{\rm rel}$ is the velocity between $J/\psi$ and a gluon where, 
\begin{equation}
v_{\rm rel}=1-\frac{{\bf k}\cdot{\bf P}}{k^0M_T\cosh y} \label{eq5}
\end{equation}
Change of variables ($K \leftrightarrow Q$) can be obtained by using Lorentz transformations:
\begin{eqnarray}
 k^0&=&(q^0E+q(\sin \theta_p\sin \theta_q\sin\phi_q+\cos\theta_p\cos\theta_q))/M_{J/\psi}\\
{\bf k}&=&{\bf q}+\frac{qE}{|{\bf P}|M_{J/\psi}}[(qM_T\cosh y-M_{J/\psi})
(\sin \theta_p\sin \theta_q\sin\phi_q+\cos\theta_p\cos\theta_q)+|{\bf P}|]
{\bf v_{J/\psi}}\label{eq7}
\end{eqnarray}
where ${\bf v_{J/\psi}}={\bf P}/E$,  $P = (E,0,|{\bf P}|\sin\theta_p,|{\bf P}|\cos\theta_p)$ and
${\bf q} = (q\sin\theta_q\cos\phi_q,q\sin\theta_q\sin\phi_q,q\cos\theta_q)$
In the rest frame of $J/\psi$, numerator of the Eq.~(\ref{eq3}) can be 
written as
\begin{equation}
\int d^3q\frac{M_{J/\psi}}{E}\sigma(q^0)f(k^0,\xi,p_{hard})\label{eq8},
\end{equation}
while, the denominator of Eq.~(\ref{eq3}) can be written as \cite{12ofplb}
\begin{eqnarray}
 \int d^3kf(k^0,\xi,p_{\rm hard})=\int d^3kf_{{\rm iso}}(\sqrt{\bf{k}^2
+\xi(\bf{k}\cdot
\hat{\bf{n}})^2},p_{\rm hard})=\frac{1}{\sqrt{1+\xi}}8\pi\zeta(3)p_{\rm hard}^3\label{eq9}
\end{eqnarray}
where $\zeta(3)$ is the Riemann zeta function. 
The maximum value of the gluon $J/\psi$ dissociation cross section \cite{KSH} is about 3 $mb$
in the range 0.7$ \leq q^0\leq$1.7 GeV. Therefore high-momentum gluons do not see the large object and simply
passes through it. On the other hand, the low-momentum gluons cannot resolve the compact object and cannot raise
the constituents to the continuum.

\subsection{Survival probability of $J/\psi$ in an anisotropic media}

To calculate the survival probability of $J/\psi$ in an anisotropic plasma 
we consider only longitudinal expansion of the matter. With the velocity 
averaged dissociation cross sections, the survival probability of the 
$J/\psi$ in the deconfined quark-gluon plasma is of the 
following form~\cite{xu},
\begin{eqnarray}
S(P_T)=\frac{\int d^2r(R^2_A-r^2)\exp[-\int^{\tau_{\rm max}}_{\tau_i}
d\tau n_g(\tau)<\sigma(K.u)v_{\rm rel}>_k]}{\int d^2r(R^2_A-r^2)}\label{eq10}
\label{spt}
\end{eqnarray}
The upper integration limit $\tau_{\rm max}=min(\tau_\psi,\tau_c)$ 
and $\tau_i$ is the QGP formation time. $n_g(\tau)$ is the gluon density at 
a given time $\tau$.  Now the $J/\psi$ will travel a distance in the transverse 
direction with velocity ${\bf v}_{J/\psi}$:
\begin{eqnarray}
d=-r\cos\phi+\sqrt{R_A^2-r^2(1-cos^2\phi)}\label{eq11}
\end{eqnarray}
Here $\cos\phi={\hat v}_{J/\psi}\cdot\hat{r}$.
The time interval $\tau_{\psi}=M_Td/P_T$ is the time before $J/\psi$ escapes from a gluon gas 
of transverse extension $R_A$. In the case of anisotropic QGP
$\tau_c$ is determined by the condition :
$p_{\rm hard}(\tau=\tau_c) = T_c$~\cite{mauricio1}, where $T_c \sim 170 - 200$ MeV. 

The modifications in Eq.(\ref{spt}) in anisotropic media come from the gluon 
density, $n_g(\tau)$ and the velocity weighted cross-section. The former is 
given by $n_g(\tau)=16\zeta(3)p_{\rm hard}^3(\tau)/[\pi^2 \sqrt{1+\xi(\tau)}]$ and
the latter has been discussed in the previous section.
The time evolution of $\xi$ and $p_{\rm hard}$ is described in the
next section.

\subsection{Space time evolution}
\subsubsection{Model I}

For an expanding plasma the anisotropy parameter $\xi$ and the hard momentum
scale $p_{\rm hard}$ (appearing in Eq.(\ref{spt})) are time
dependent.
Thus to calculate $S(P_T)$
one needs to know the time dependence of $p_{\rm hard}$ and $\xi$.
To obtain the time evolution of the parameters
we shall follow the work of Ref.~\cite{mauricio1,LP4} and evaluate 
$S(P_T)$ of the $J/\psi$
 from the first few fermi of the 
plasma evolution. 

\begin{figure}
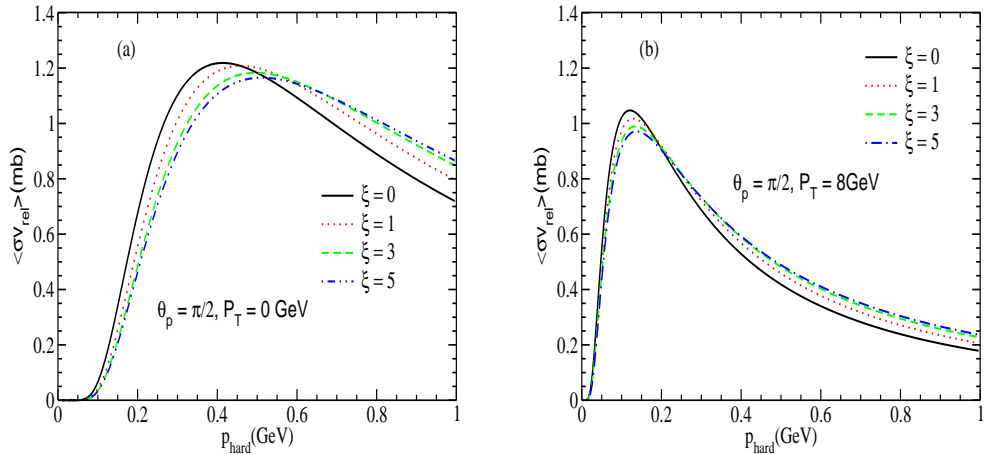

\begin{center}
\epsfig{file=cpt0.eps,width=6cm,height=6cm,angle=0}~~~~~~~\epsfig{file=cpt8.eps,width=6cm,height=6cm,angle=0}
\caption{(Color online) 
The thermal-averaged gluon-$J/\psi$ dissociation cross section as function
of the hard momentum scale at central rapidity ($\theta_p=\pi/2$) for
$\xi = \{0, 1, 3, 5\}$. 
(a) corresponds to $P_T = 0$
and (b) is for $P_T= 8$ GeV.
}
\label{fig1}
\end{center}
\end{figure}
The 
time dependence of relevant quantities is given by~\cite{mauricio1},
\begin{eqnarray}
\xi(\tau,\delta) &=& \left(\frac{\tau}{\tau_i}\right)^
{\delta(1-\Lambda(\tau))}-1,\nonumber\\
p_{\rm hard}(\tau)&=&T_i~{\bar {\cal U}}^{c_s^2}(\tau),
\label{eq14}
\end{eqnarray}
where, 
\begin{eqnarray}
{\mathcal U}(\tau)&\equiv& \left[{\mathcal
    R}\left((\frac{\tau_{\rm iso}}{\tau})^\delta-1\right)
\right]^{3\Lambda(\tau)/4}
\left(\frac{\tau_{\rm iso}}{\tau}\right)^{1-\delta(1-\Lambda(\tau))/2},
\nonumber\\
{\bar {\cal U}}&\equiv& \frac{{\cal U}(\tau)}{{\cal U}(\tau_i)},
\nonumber\\
{\mathcal R}(x)&=&\frac{1}{2}[1/(x+1)+\tan^{-1}{\sqrt
    {x}}/\sqrt{x}]
\label{eq15}
\end{eqnarray}
where the exponent $\delta = 2$ corresponds to  free-streaming 
pre-equilibrium momentum space anisotropy and
$\delta=0$ corresponds to thermal equilibrium, 
$T_i$ is the initial temperature of the plasma
and $c_s^2$ is the velocity of
sound. 
$\Lambda(\tau)=\frac{1}{2}(\tanh[\gamma(\tau-
\tau_{\rm iso})/\tau_{\rm iso}]+1)$ is the smeared step function
 introduced to take into account 
the smooth transition from non-zero value of $\delta$ to $\delta=0$
at $\tau=\tau_{\rm iso}$~\cite{dilstrick} with $\gamma$ being the transition
width. 
\begin{figure}
\begin{center}
\epsfig{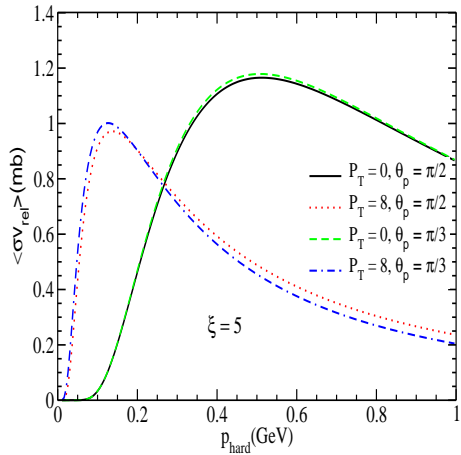}
\caption{(Color online) Direction dependence of thermal averaged 
gluon-$J/\psi$ dissociation cross section for two values of $P_T$ and
at two different rapidities.}  
\label{fig2}
\end{center}
\end{figure}

 For isotropic case, we have $p_{\rm hard} = T, \tau_{\rm iso}=\tau_i$ 
so that $\Lambda = 1$, ${\cal U}(\tau)=\tau_i/\tau$, and ${\cal U}(\tau_i)=1$. 
By using $c_s^2 =1/3$ we recover
the Bjorken cooling law~\cite{bj}.
As the colliding nuclei do have a transverse density profile,
we assume that the initial temperature profile is given by~\cite{moore}
\begin{eqnarray}
T_i(r) = T_i\,\left[2\left(1-r^2/R_A^2\right)\right]^{1/4}
\label{eq16}
\end{eqnarray}
Using Eqs.(\ref{eq14}) and (\ref{eq16}) we obtain the profile
of the hard momentum scale as
\begin{eqnarray}
 p_{\rm hard}(\tau,r) = T_i\,\left[2\left(1-r^2/R_A^2\right)\right]^{1/4} 
{\bar {\cal U}}^{c_s^2}(\tau)\label{eq17}
\end{eqnarray}
We use two sets of initial conditions for RHIC energies. The first set,
henceforth referred to as Set I, corresponds to that used in Ref.~\cite{xu} , 
i.e., $T_i = 550$ MeV and $\tau_i = 0.7$ fm/c. For other set (set II)
the initial temperature (time) has been calculated using the measured
multiplicities at RHIC energies~\cite{mahatsab} and is given by
$T_i = 440$ MeV corresponding to $\tau_i = 0.15$ fm/c. 
At LHC energies we use the initial conditions:
$T_i = 820 $ MeV and $\tau_i = 0.5 $ fm/c. 

\subsubsection{Model II}
The other alternative scenario of time dependence for $\xi$ and
$p_{\rm hard}$ in highly anisotropic system has been described 
in ~\cite{martinez2010} taking
the first two moments of Boltzmann equation which reads in (0+1)-dimension
as
\begin{equation}
E\frac{\partial f(t,z,{\bf p})}{\partial t}+
p_z\frac{\partial f(t,z,{\bf p})}{\partial z} 
=-{\cal C}[f(t,z,{\bf p})]
\end{equation}
Without going into further details we simply quote the coupled differential
equations that has to be solved to get the time dependence of
$\xi$ and $p_{\rm hard}$~\cite{martinez2010}:
\begin{equation}
\frac{1}{1+\xi}\partial_{\tau}\xi=\frac{2}{\tau}-4\Gamma{\cal R}(\xi)\,
\frac{{\cal R}^{3/4}\sqrt{1+\xi}-1}{2{\cal R}(\xi)
+3(1+\xi){\cal R}^\prime(\xi)}
\end{equation}
\begin{equation}
\frac{1}{1+\xi}\frac{1}{p_{\rm hard}}\partial_{\tau}p_{\rm hard}=
\Gamma{\cal R}^\prime(\xi)\,
\frac{{\cal R}^{3/4}\sqrt{1+\xi}-1}{2{\cal R}(\xi)
+3(1+\xi){\cal R}^\prime(\xi)}
\end{equation}
where $\Gamma = 2T(\tau)/(5{\bar \eta})$ and ${\bar \eta}=\eta/s$, $\eta$
is the shear viscosity co-efficient. 
In this model the time $\tau_c$ has been calculated using the relation
${\cal R}^{1/4}(\xi)\,p_{\rm hard} = T_c$. We have used the same transverse
profile for the hard momentum scale as in model I.
A comparative study
of the survival probability using the above described space-time 
models will be done.

\begin{figure}
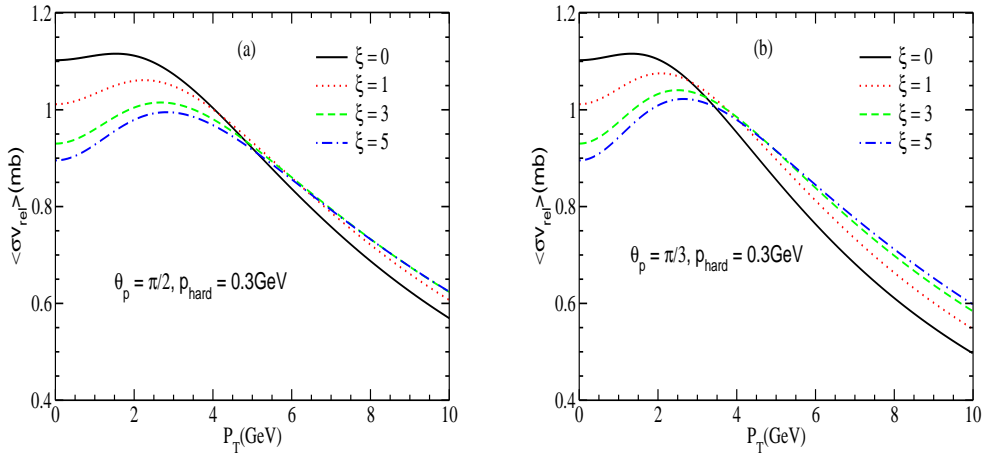

\begin{center}
\epsfig{file=cT0.3.eps,width=6cm,height=6cm,angle=0}~~~~~~~\epsfig{file=f2T0.3.eps,width=6cm,height=6cm,angle=0}
\caption{(Color online) 
The thermal-averaged gluon-$J/\psi$ dissociation cross section 
as a function of the transverse momentum $P_T$ for $p_{\rm hard}$
at (a) central and (b) forward rapidity regions. 
}
\label{fig3}
\end{center}
\end{figure}
\section{Results}

Let us first discuss the thermal averaged gluon dissociation cross section
in anisotropic system. Eqs.(4)-(9) have been used for this purpose. The
results are displayed in Fig.~(\ref{fig1}). Fig.(1a) and Fig.(1b) correspond to 
$P_T$=0 and $P_T$=8 GeV respectively for a set of values of the anisotropy parameter. It
is seen that the cross section decreases with $\xi$ for $p_{\rm hard}$
up to $\sim$ 500 MeV and then increases as compared to the isotropic case
($\xi=0$)(see in Fig.(1a)). 
Similar feature has been observed  in Fig.(1b) for higher $P_T$ 
where the cross section starts to increase beyond $p_{\rm hard} \sim 200$ MeV. 

\begin{figure}
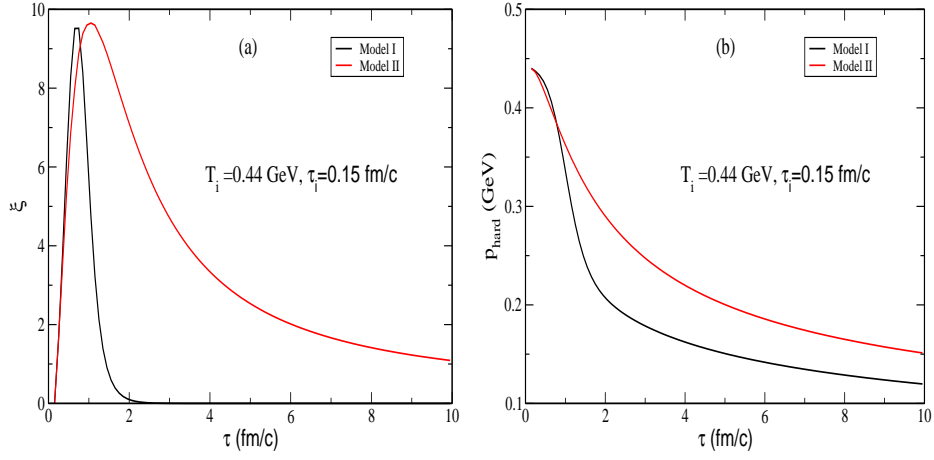

\begin{center}
\epsfig{file=xi_I_II.eps,width=6cm,height=6cm,angle=0}~~\epsfig{file=phard_I_II.eps,width=6cm,height=6cm,angle=0}
\caption{(Color online) 
Time evolutions of (a) the anisotropy parameter $\xi$
and (b) the hard momentum scale
$p_{\rm hard}$  in the two space time models described in the 
text.
}
\label{fig4a}
\end{center}
\end{figure}

\begin{figure}
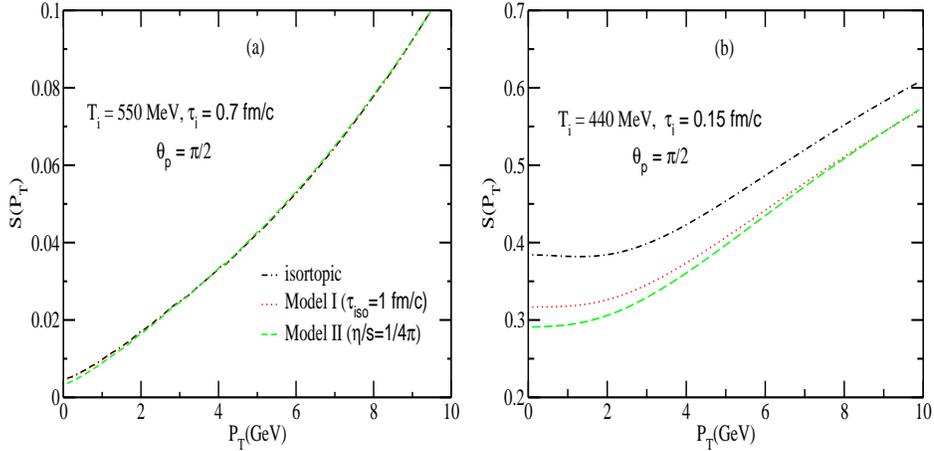

\begin{center}
\epsfig{file=RHIC1a.eps,width=6cm,height=6cm,angle=0}~~\epsfig{file=RHIC1b.eps,width=6cm,height=6cm,angle=0}
\caption{(Color online) The survival probability of $J/\psi$ in an anisotropic plasma at central rapidity.  
(a) corresponds to $T_i$=550 MeV, $\tau_i$=0.7 fm/c and (b) is for $T_i$=440 MeV, $\tau_i$=0.15 fm/c.
}
\label{fig4}
\end{center}
\end{figure}

As mentioned before, the dissociation cross section depends on the
direction of propagation ($\theta_p$) of the quarkonium with respect to 
the anisotropy axis. This dependence is shown in Fig.~(\ref{fig2}). We
find marginal dependence in this case. These observations will have important
consequences while calculating the survival probability (see later).

In order to show the transverse momentum dependence for fixed hard momentum
scale, we, in Fig.~(\ref{fig3}), present the dissociation cross section
as a function of $P_T$ of the $J/\psi$ for $p_{\rm hard} = 300$ MeV and
for two values of $\theta_p$. Again it is seen 
that the cross section first decreases with the anisotropy parameter
upto $P_T \sim 5 (3)$ GeV for $\theta_p  =\pi/2 (\pi/3)$ and we find
larger increase away from the central rapidity region. This
behavior might influence the survival probability
which we consider next.

Before calculating the survival probability let us examine the time
evolutions of $\xi$ and $p_{\rm hard}$ in the space time models described
earlier. This is needed to calculate the survival probability in
an expanding plasma. The results are shown in Fig.~(\ref{fig4a}).
It is seen that the anisotropy parameter falls much rapidly compared to
the case when model II is used (see Fig.(4a)). There is a narrow window in $\tau$ where
$\xi$ dominates in case of model I. The cooling is slower in case of model II as
can be seen from Fig.(4b). These observations
have important consequence on the survival probability as we shall see. 

Eq.(\ref{spt}) has been used to calculate the survival probability.
For the space time model I we use Eq.(14) and for model II Eqs.(19) and
(20) have been used for the time evolution of the anisotropy parameter
$\xi$ and the hard momentum scale $p_{\rm hard}$.
Fig.(\ref{fig4}) describes the survival probability
for two different set (Set I and Set II) of initial conditions for a given direction of
propagation of the $J/\psi$ with the anisotropy axis. 
For the Set I initial conditions, the results
are same for the isotropic case and the two space time models
used for the anisotropic media (see in Fig.(5a)).
However, for the Set II initial conditions and for the
same $\theta_p$, the results are different from each other as can be seen from Fig.(5b). 
More interestingly, we find an order of magnitude 
increase in the survival probability for the Set II initial conditions.
This is because of the argument of the exponential in Eq.(10).

Next we consider the survival probability in forward rapidity
region for the two sets of initial conditions and two space time
models and compare it with that in the central rapidity region. 
Fig.(6a)((6b)) shows the survival
probability calculated using Set I (Set II) initial conditions.
It is seen that for the Set I initial conditions,
$S(p_T)$, in the forward rapidity region is marginally larger than the 
case when
space time model II is used. It is seen from Fig.(6b) that for Set II initial conditions, 
$S(p_T)$ is marginally higher
in the case of space time model I in the low $p_T$ region at central
rapidity. However,
at forward rapidity, the survival probability is always larger
in case of space time model I.

In order to show the dependences of $S(p_T)$ on $\tau_{\rm iso}$
in the space time model I and on $\eta/s$ in space time model II
we plot the survival probability for set I initial conditions
in Fig.(\ref{fig6}). It is observed that increasing $\tau_{\rm iso}$
lowers $S(p_T)$ while increasing $\eta/s$ enhances the survival probability.

\begin{figure}
\begin{center}
\epsfig{file=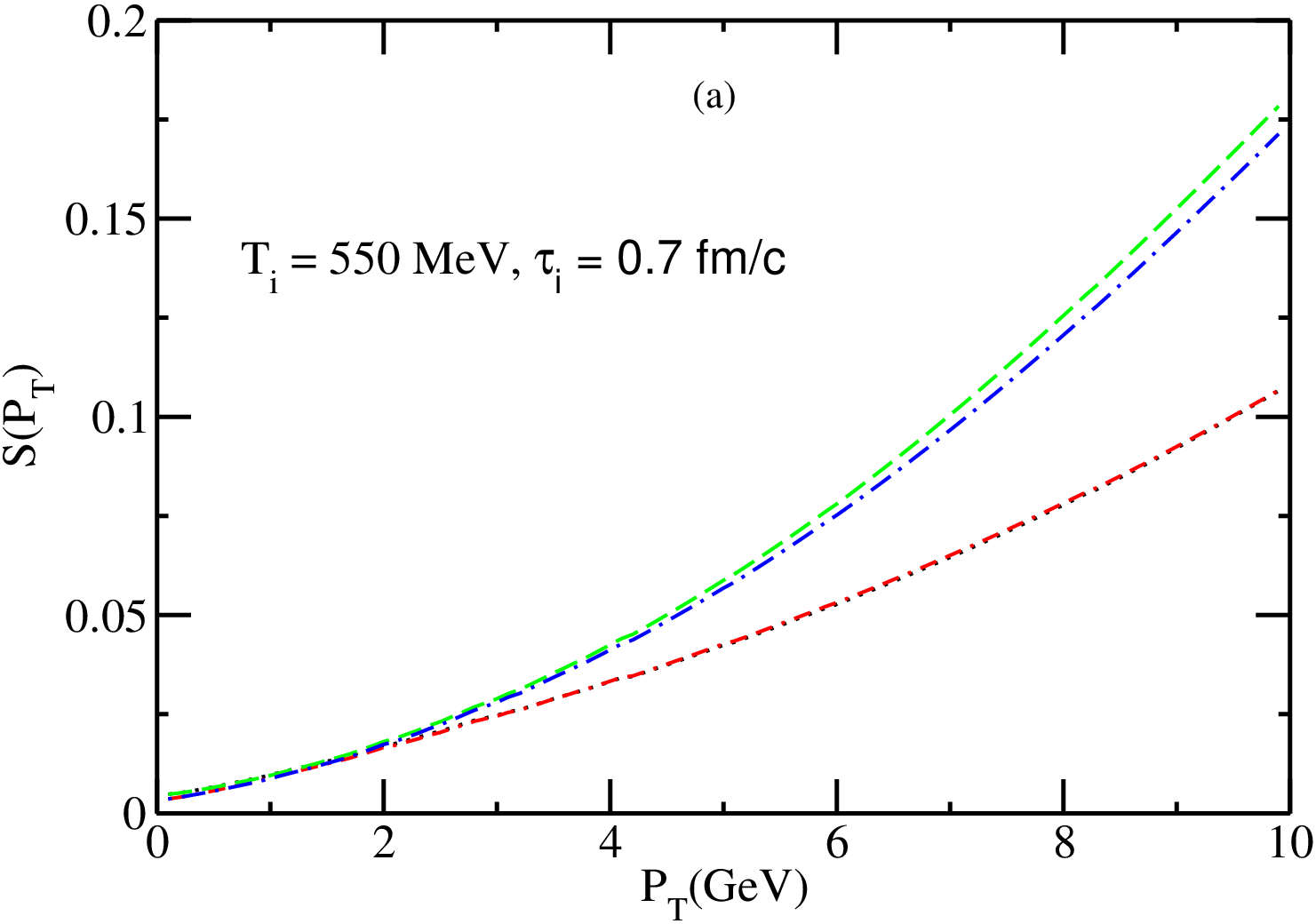,width=6cm,height=6cm,angle=0}~~\epsfig{file=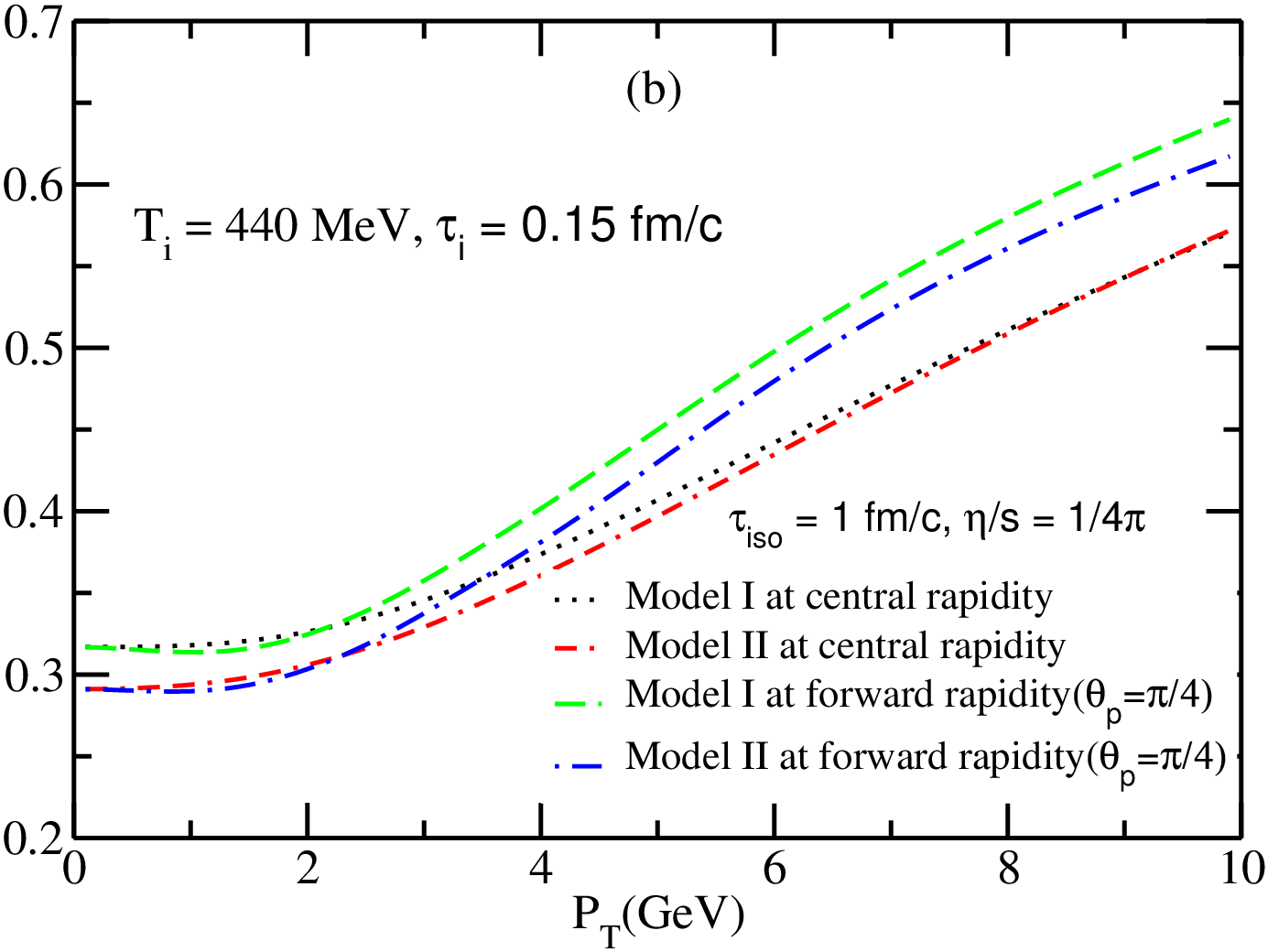,width=6cm,height=6cm,angle=0}
\caption{(Color online) The survival probability of $J/\psi$ in an 
anisotropic plasma at central and forward rapidity regions 
corresponding to (a) $T_i$=550MeV, $\tau_i$=0.7 fm/c and (b) $T_i$=440 MeV, $\tau_i$=0.15fm/c.} 

\label{fig5}
\end{center}
\end{figure}


The results for the LHC energies for the two sets of space time
models are shown in Fig.~(\ref{fig7}). In case of central rapidity region
we do not find any difference in the results for the two space time models.
But the results are marginally different in case of forward rapidity region.
However, in the forward rapidity region we observe that the survival 
probability increases by a factor of 2 compared to the case of
central rapidity for a given set of initial conditions.

\begin{figure}
\begin{center}
\epsfig{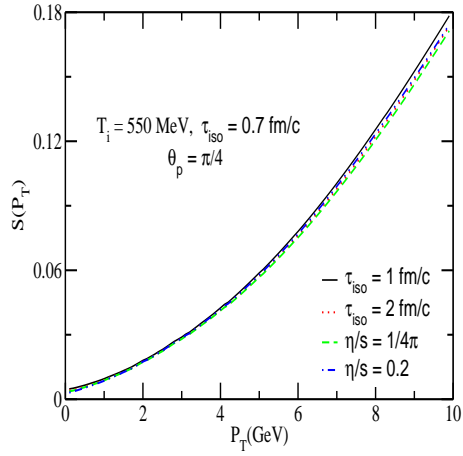}
\caption{(Color online) The survival probability of $J/\psi$ in an 
anisotropic plasma for two different model at same initial temperature 
$T_i=500$ MeV and initial time $\tau_i=0.7$ fm/c for different $\tau_{\rm iso}$
and $\eta/s$.
}
\label{fig6}
\end{center}
\end{figure}

\begin{figure}
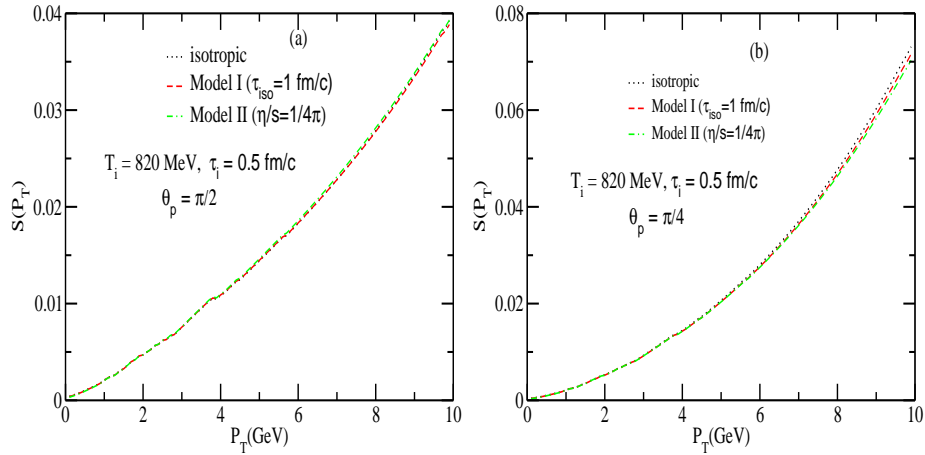

\begin{center}
\epsfig{file=lhcc.eps,width=6cm,height=6cm,angle=0}~\epsfig{file=lhcf.eps,width=6cm,height=6cm,angle=0}
\caption{(Color online) The survival probability of $J/\psi$ in an 
anisotropic plasma for (a) central and (b) forward rapidity region. 
The initial conditions are taken $T_i=820$ Mev and $\tau_i=0.5$ fm/c. 
}
\label{fig7}
\end{center}
\end{figure}


\section{Summary}

We have calculated gluon-$J/\psi$ dissociation cross section assuming
pre-equilibrium momentum space anisotropy in the deconfined phase
expected to be produced in relativistic heavy ion collisions.
It is observed that the thermally weighted cross section is modified
substantially in anisotropic plasma. To calculate the survival
probability of the $J/\psi$ two sets of initial conditions and
two different space models have
been used both for RHIC and LHC energies. 
For set I initial conditions, in the central rapidity region
we do not find any difference in the survival probabilities calculated
in the isotropic and anisotropic QGP at RHIC energies. However, changing 
the initial conditions, it is seen that the survival probability
is lower in AQGP. Moreover, there is noticeable difference in the
results obtained using different space time models.
We also show that the results for the survival probability
depend on the isotropization time (in model I) and $\eta/s$ (in
model II).
For the case of LHC with a given initial conditions the
results marginally differ from each other in the forward
rapidity region when two different space time models are
used.
It is also found that the results are not much sensitive
to the direction of propagation of the $J/\psi$ with respect to
the anisotropy axis. 
It is also demonstrated that
the results are extremely sensitive to the initial conditions,
in particular, to the choice of the initial time. It is important to note 
that uncertainty may result from our assumption of chemical equilibrium.
However, one can naively expect that finite chemical potentials
should affect isotropic and anisotropic plasmas equally. So one expects 
that although the total yields could change, one would still see 
a sensitivity to the assumed isotropization or thermalization time.
At leading order in the quark fugacity,
the ratio of the isotropic to anisotropic result should be
independent of the fugacity~\cite{plb678, plb331}. 
We would also like
to add that the consideration of transverse expansion may
alter the results during the late stages of the collisions as
has been observed in case of photon and dilepton transverse momentum
distribution. However, transverse expansion is pronounced in the
later stage and its effect in the very early stage is minimal. Since
in our case momentum anisotropy is an early stage phenomena, the
effect will be negligible. 
We also not that treating the quarkonium binding
energy as function of $\xi$ and $p_{\rm hard}$ might alter the present
findings and this is worth investigating.

It is to be noted that apart from this mechanism of suppression, there are 
other mechanisms by which $J/\psi$ can be suppressed~\cite{xu}. All these
possible processes should be taken into account and then be compared with
the experimental data of transverse momentum distribution of $J/\psi$
to extract the isotropization time $\tau_{\rm iso}$ as has been
done in case of photons~\cite{LB3} and nuclear modification
of light hadrons~\cite{mahatsab}.

\end{document}